\documentstyle[aps,epsf,preprint,prb]{revtex}
\def\ba{\begin{array}}
\def\ea{\end{array}}
\def\bc{\begin{center}}
\def\ec{\end{center}}
\def\be{\begin{equation}}
\def\ee{\end{equation}}
\def\bnn{\begin{eqnarray*}}
\def\enn{\end{eqnarray*}}
\def\bi{\begin{itemize}}
\def\ei{\end{itemize}}
\def\bn{\begin{eqnarray}}
\def\en{\end{eqnarray}}
\def\bt{\begin{tabular}}
\def\et{\end{tabular}}
\def\bdes{\begin{description}}
\def\edes{\end{description}}

\def\ds{\displaystyle}

\def\nc{\nonumber \\}

\def\a{\alpha}
\def\b{\beta}
\def\d{\delta}
\def\dkot{{\delta_{\vko,\vkt}}}

\def\f{\frac}
\def\g{\gamma}

\def\o{\omega}

\def\s{\sigma}

\def\O{\Omega}
\def\Ob{{\Omega^b}}
\def\Okb{{\Omega_\vk^b}}
\def\Omkb{{\Omega_{-\vk}^b}}
\def\Okob{{\Omega_{\vk_1}^b}}

\def\Oktb{{\Omega_{\vk_2}^b}}

\def\Okc{{\Omega_\vk^c}}
\def\Okoc{{\Omega_{\vk_1}^c}}
\def\Oktc{{\Omega_{\vk_2}^c}}
\def\Oqc{{\Omega_\vq^c}}

\def\Okd{{\Omega_\vk^d}}
\def\Okod{{\Omega_{\vk_1}^d}}
\def\Oktd{{\Omega_{\vk_2}^d}}

\def\Opb{{\Omega_\vp^c}}

\def\Oqd{{\Omega_\vq^d}}

\def\tu{{\tilde{u}}}

\def\va		{{\bf a}}

\def\vi		{{\bf i}}
\def\vj		{{\bf j}}
\def\vk		{{\bf k}}
\def\vko	{{{\bf k}_1}}
\def\vkt	{{{\bf k}_2}}
\def\vl		{{\bf l}}

\def\vp		{{\bf p}}
\def\vq		{{\bf q}}

\def\vx		{{\bf x}}
\def\vy		{{\bf y}}

\def\vQ		{{\bf Q}}

\def\ua{u^\ast}

\def\pkc{{\phi_{\vk}^c}}
\def\pkmc{{\phi_{-\vk}^c}}

\def\pkzc{{\phi_{\vk 0}^c}}

\def\pkd{{\phi_{\vk}^d}}
\def\pkmd{{\phi_{-\vk}^d}}

\def\pkzd{{\phi_{\vk 0}^d}}

\def\pqzc{{\phi_{\vq 0}^c}}
\def\pqzd{{\phi_{\vq 0}^d}}
\def\dag{\dagger}

\def\lg{\langle}
\def\rg{\rangle}

\def\ra{\rightarrow}

\def\ua{\uparrow}
\def\da{\downarrow}

\def\svk{\sum_{\vk}}

\def\I0{{\bm{cc}1&0\\0&1\em}}

\begin{document}
\draft
\title{
Functional Schr\"odinger Representations
of Holstein-Primakoff Boson and Slave-Boson Theories
for Heisenberg Antiferromagnets}
\author{Sul-Ah Ahn, Kwangyl Park and Sung-Ho Suck Salk}
\address{Department of Physics, 
Pohang University of Science and Technology,
Pohang 790-784, Korea} 
\date{\today}
\maketitle

\begin{abstract}
We present  
functional Schr\"odinger representations
of Holstein-Primakoff boson 
and slave-boson theories
for the Heisenberg Hamiltonian and 
the $t-J$ Hamiltonian respectively.
Based on these representations
we obtain the dispersion relations of magnons for
two dimensional  
antiferromagnets.
By applying 
the functional Schr\"odinger representation
of the Holstein-Primakoff boson theory
to the Heisenberg Hamiltonian,
the exchange energy 
is correctly predicted 
and the self-energy 
of quasi-hole
is obtained.
From the use of
the functional Schr\"odinger representation
of the $t-J$ Hamiltonian it is shown that
at half-filling
the dispersion relation 
obtained from the slave-boson theory
leads to that
obtained from the Holstein-Primakoff boson approach.
\end{abstract}

\vspace{1cm}

\section{INTRODUCTION}
Although widely used 
in the area of high energy physics,
the functional Schr\"odinger picture theory
\cite{fj,Hat,Barnes,ss,kw,jk,dm,ej,ky}
is relatively new
in condensed matter physics.
Lately, only a limited number of application 
to condensed matter physics
appeared in the literature.\cite{nyk,nkn,Ahn}
Earlier we treated the $t-J$ Hamiltonian 
by introducing a slave-boson approach 
to the functional Schr\"odinger picture(FSP) theory\cite{fj} 
for the two-dimensional systems
of antiferromagnetically correlated systems\cite{Ahn}.
The Holstein-Primakoff boson representation
is often used to describe
the broken symmetry phases
of the quantum Heisenberg antiferromagnet.
Lately Chang\cite{Chang} introduced
a generalized Holstein-Primakoff representation
of the $t-J$ model Hamiltonian
in order to describe a higher order(second order) effect
on spin waves and antiferromagnetic spin polarons
by allowing a systematic perturbative expansion.
Here we present
the functional Schr\"odinger representations
of both the Holstein-Primakoff\cite{Chang} boson
and slave-boson\cite{Ahn} theories
for the Heisenberg Hamiltonian 
and the $t-J$ Hamiltonian
respectively.
Based on these representations
we derive magnon dispersion relations
and present comparison 
between the two approaches.
\section{
Holstein-Primakoff Boson Theory
of Heisenberg Hamiltonian
by Functional Schr\"odinger Representation;
Antiferromagnetic Magnon
} 
To allow symbol definitions
for later use,
here we choose a brief review 
on a generalized approach of
perturbatively treated Heisenberg Hamiltonian
based on a Holstein-Primakoff transformation
\cite{Chang}.
The Hilbert space of present interest 
is spanned by the states, 
$|a_1 \rg \otimes |a_2\rg\otimes\cdots\otimes 
|a_j \rg\otimes\cdots\otimes
 |a_N \rg$,
where $|a_j \rg \in \{c^\dag _{j\uparrow}|0\rg _j),
c^{\dag }_{j\downarrow}|0 \rg_j, |0\rg_j \}$
with $c_{j\s}|0\rg_j=0$.
The vacuum state is given by $|0\rg=\otimes|0\rg_j$ 
which satisfies $c_{j\s}|0\rg =0$.
$j$ is the site index
and $\s$, the spin index. 
$N$ is the total number of lattice sites.
With the local Hubbard operators $X^{ab}_j=|a_j\rg \lg b_j|$,
the Heisenberg Hamiltonian
for the two-dimensional system 
of antiferromagnetically correlated electrons 
is written\cite{Onu} 
\be
H_J=\f{J}{2}\sum_{\lg i, j \rg}\left(X_{i}^{\s -\s}X^{-\s\s}_{j}
-X_{i}^{\s\s}X_{j}^{-\s-\s}\right),
\label{hs0}
\ee
where $J$ is the  Heisenberg coupling constant.
$\lg ij\rg$ stands for 
summation only over
nearest neighbours. 
The above Hamiltonian
includes the contribution of 
the $\f{1}{4}n_in_j$ term that appears 
in the usual $t-J$ Hamiltonian,
where $n_i=\sum_\s c_{i\s}^\dag c_{i\s}$.
Introducing two commuting 
boson(magnon) operators $(a_j,b_j)$
and an anticommuting 
fermion(hole) operator $f_j$ 
for a bipartite lattice 
made of sublattices, $A$ and $B$,
two sets of Holstein-Primakoff representations\cite{Chang} 
for the local Hubbard operators are given
in the table below.
\[\begin{array}{|c||c|c|}\hline
Hubbard ~~operator & A~~sublattice & B~~sublattice \\ \hline
X_j^{00} & f_j^\dag f_j & f_j^\dag f_j\\
X_j^{0\ua} & f_j^\dag a_j 
&\sqrt{2s}f_j^\dag \sqrt{1-\f{1}{2s}\left(b_j^\dag b_j
+f_j^\dag f_j\right)}\\
X_j^{\ua 0} &a_j^\dag f_j 
&\sqrt{2s}\sqrt{1-\f{1}{2s}\left(b_j^\dag b_j
+f_j^\dag f_j\right)}f_j\\
X_j^{0\da}
& \sqrt{2s}f_j^\dag \sqrt{1-\f{1}{2s}\left(a^\dag_j a_j
+f_j^\dag f_j\right)} 
& f_j^\dag b_j \\
X_j^{\da 0} 
& \sqrt{2s}\sqrt{1-\f{1}{2s}\left(a^\dag_j a_j
+f_j^\dag f_j\right)}f_j 
& b_j^\dag f_j \\
X_j^{\ua\ua}& a_j^\dag a_j  
& 2s\left[1-\f{1}{2s}\left(b_j^\dag b_j
+f_j^\dag f_j\right)\right] \\
X_j^{\da\da} 
& 2s\left[1-\f{1}{2s}\left(a_j^\dag a_j
+f_j^\dag f_j\right)\right] 
& b_j^\dag b_j \\
X_j^{\ua\da} 
& \sqrt{2s}a_j^\dag\sqrt{1-\f{1}{2s}
\left(a_j^\dag a_j+f_j^\dag f_j\right)}
& \sqrt{2s}\sqrt{1-\f{1}{2s}\left(b_j^\dag b_j
+f_j^\dag f_j\right)}b_j\\ \hline 
\end{array}
\]
For sublattice $A$,
$a_j(a_j^\dag)$ is the annihilation(creation) operator
of spin-up boson (magnon excitation)
and $f_{j}(f_j^\dag)$, the annihilation (creation) operator
of spinless fermion (hole excitation) at site j.
$s$ is the spin quantum number.
Likewise, for sublattice $B$, $b_j(b_j^\dag)$ 
is the annihilation(creation) operator 
of spin-down boson(magnon excitation)
and $f_j(f_j^\dag)$, the annihilation(creation) operator 
of spinless fermion(hole excitation) at site j.

Using the local Hubbard operators\cite{Chang} 
in the table above, 
we derive from Eq.(\ref{hs0}),
\bn
H_J&=& J\sum_{\lg i,j\rg}\left\{s a_i^\dag\sqrt{1-
\f{1}{2s}\left(a_i^\dag a_i+f_i^\dag f_i\right)}
b_j^\dag
\sqrt{1-\f{1}{2s}\left(b_j^\dag b_j
+f_j^\dag f_j\right)}\right. \nc
&&-\f{1}{2}a_i^\dag a_i b_j^\dag b_j
+s\sqrt{1-\f{1}{2s}\left(a_i^\dag a_i+f_i^\dag f_i\right)
}a_i\sqrt{1-\f{1}{2}\left(b_j^\dag b_j+f_j^\dag f_j\right)}
b_j \nc
&&\left.-2s^{2}\left[1-\f{1}{2s}
\left(a_i^\dag a_i + f_i^\dag f_i\right)\right]
\left[1-\f{1}{2s}
\left(b_j^\dag b_j+f_j^\dag f_j\right)\right]\right\}
\label{hj}
\en
The transformed Heisenberg Hamiltonian above 
is perturbatively treated
by using the following expansions;
\bn
\sqrt{2s}\sqrt{1-\f{1}{2s}\left(a^\dag_i a_i
+f_i^\dag f_i\right)}&=&
1-\f{1}{4s}\left(a^\dag_i a_i+f_i^\dag f_i\right)
+ (H.O.),\nc
\sqrt{2s}\sqrt{1-\f{1}{2s}\left(b^\dag_j b_j
+f_j^\dag f_j\right)}&=& 
1-\f{1}{4s}\left(b^\dag_j b_j+f_j^\dag f_j\right)
+(H.O.). 
\label{pe}
\en

We consider terms up to order $\f{1}{s}$ in Eq. (\ref{pe})
for the evaluation of Eq. (\ref{hj}).
By first taking the Fourier transformation of 
Eq. (\ref{hj})
and then the Bogoliubov transformation,
we obtain the following four terms;
1) the free Hamiltonian $H_{0}$,
\bn
H_{0}&=&\svk\o_{\vk}^0\left[1-\f{1}{2s}\f{2}{N}\sum_{\vq}
\left(\f{\o_\vq^0}{Jsz}-1\right)\right]
\left(c_{\vk}^{\dag}c_{\vk}
+d_{\vk}^{\dag}d_{\vk}\right)
+Jsz\svk\left(\f{\o_\vk^0}{Jsz}-1
-2f_\vk^\dag f_\vk\right)\nc
&&
-Js^2zN,
\label{hz}
\en
with $N$, the total number of spins.
2) The magnon-magnon interaction, $H_{M-M}$,
\bn
H_{M-M}&=&-\f{Jz}{4}\f{2}{N}\sum_{\vko,\vkt}\left[
\left(1-\f{\o_\vko^0
\o_\vkt^0}{(Jsz)^2}\right)
\left(c_\vko^{\dag}c_\vko c_\vkt^{\dag}c_\vkt
+d_\vko^{\dag}d_\vko d_\vkt^{\dag}d_\vkt\right)\right.\nc
&&\left.+2\left(1+\f{\o_\vko^0\o_\vkt^0}{(Jsz)^2}\right)
 c_\vko^{\dag}c_\vko d_\vkt^{\dag}d_\vkt\right],
\label{hmm}
\en
3) the magnon-hole interaction, $H_{M-H}$,
\be
H_{M-H}=-\f{Jz}{2}\f{2}{N}\sum_{\vko,\vkt}
\f{\o_\vkt^0}{Jsz}f_\vko^\dag f_\vko
\left(c_\vkt^\dag c_\vkt
+d_\vkt^\dag d_\vkt\right),
\label{hmh}
\ee
and 4)the hole-hole interaction, $H_{H-H}$,
\be
H_{H-H}=-\f{Jz}{2}\f{2}{N}\sum_{\vko,\vkt,\vq}
\g_{\vq}f_{\vko-\vq}^\dag f_\vko f_{\vkt+\vq}^\dag f_\vkt,
\label{hhh}
\ee
where $\o_\vk^0=Jsz\sqrt{1-\g_\vk^2}$,
$\g_\vk\equiv\f{1}{z}\sum_{\mathbf{\d}}e^{i\vk\cdot\mathbf{\d}}
=\g_{-\vk}$
with $z$, the number of nearest neighbors.
For a square lattice
we have
$\g_\vk = \f{1}{2}\left[\cos (k_xa)+\cos (k_ya)\right]$.
The last two terms in Eq. (\ref{hz})
are contributed from 
the $\f{1}{4}n_in_j$ term
that appears in the original $t-J$ Hamiltonian.
This term was not considered 
in the work of Chang\cite{Chang}.
$c_\vk(c_\vk^\dag)$ and $d_\vk(d_\vk^\dag)$
are the annihilation(creation) operators
of the chargeless bosons(quasimagnons) 
of momentum $\vk$,
in association with 
two sublattices $A$ and $B$ respectively.
$f_\vk(f_\vk^\dag)$ 
is the annihilation(creation) operator
of a spinless hole(holon) of momentum $\vk$.

For the application of 
the Holstein-Primakoff theory 
to functional  Schr\"{o}dinger picture(FSP) representation, 
one has to choose the FSP operator
of the charged fermion.
On the other hand
the charge-neutral fermion FSP operator 
is used in the slave-boson theory\cite{Ahn}.
In the following, we derive
the ground state energy 
and the dispersion energy 
of magnon for  
the two-dimensional systems 
of antiferromagnetically correlated electrons 
at and near 
half filling.

Similarly to the case of
relativistic FSP theory\cite{Hat}
we define, 
for chargeless scalar bosons(magnons),
$$
\begin{array}{ll}
~~~~~~~~~~~~~~~~~~~~~~
c_\vk=\f{1}{\sqrt{2}}\left(\pkc+\f{\d}{\d\pkc^*}\right),&
c_\vk^\dag=\f{1}{\sqrt{2}}\left(\pkc^*-\f{\d}{\d\pkc}\right),
~~~~~~~~~~~~~~~~~~~~~~~~~~~~~~~(8.a)\\ 
~~~~~~~~~~~~~~~~~~~~~~
d_\vk=\f{1}{\sqrt{2}}\left(\pkd+\f{\d}{\d\pkd^*}\right),&
d_\vk^\dag=\f{1}{\sqrt{2}}\left(\pkd^*-\f{\d}{\d\pkd}\right),
~~~~~~~~~~~~~~~~~~~~~~~~~~~~~~~(8.b)
\end{array}
\label{cd}
$$
with $\phi_\vk$, the scalar field variable 
in momentum space,
and for charged fermions(holes),
$$
\begin{array}{ll}
~~~~~~~~~~~~~~~~~~~~~~~~
f_\vk=\f{1}{\sqrt{2}}\left(u_\vk + \f{\d}{\d u_\vk^\dag}\right),&
f_\vk^\dag=\f{1}{\sqrt{2}}\left(u_\vk^\dag + \f{\d}{\d u_\vk}\right).
~~~~~~~~~~~~~~~~~~~~~~~~~~~~~~~~(8.c)
\end{array}
\label{f}
$$
with $u_\vk$, the Grassmann field variable
in momentum space.
The field variables $\phi$ and $u$ defined above 
satisfy
the following commutation and anticommutation relations
(see Appendix \ref{app1}),
\setcounter{equation}{8}
\be
\left[c_{\vk}, c_{\vk^{`}}^{\dag}\right]=\d_{\vk,\vk^{`}}, 
\left[d_{\vk}, d_{\vk^{`}}^{\dag}\right]=\d_{\vk,\vk^{`}},
\left\{f_{\vk}, f_{\vk^{`}}^{\dag}\right\}=\d_{\vk,\vk^{`}}.
\label{cr}
\ee

The functional Schr\"odinger equation
is written,
\be
H[\phi^c,\phi^d,u]\Psi_\O[\phi^c,\phi^d,u]
=E\Psi_\O[\phi^c,\phi^d,u]
\label{ge}
\ee
where $\Psi_\O$ is 
the ground state
for the square lattice 
of antiferromagnetic spin order
at or near half-filling,
\be
\Psi_\O[\phi^c,\phi^d,u]
=N_\O\exp\left[-\f{1}{2}
\svk\left(\pkmc\Okc\pkc
+\pkmd\Okd\pkd\right)
+\f{1}{2}\sum_{\vk,\vk '}u_{\vk}^{\dag}\O_{\vk,\vk^{'}}^f
u_{\vk^{'}}\right]
\label{gtf}
\ee
where $N_\O$ is a proper normalization constant.
By taking the variation of the ground state energy
with respect to the Gaussian exponents
$\O^c$, $\O^d$ and $\O^f$,
\be
E=\f{\lg\O|H|\O\rg}{\lg\O|\O\rg}=E[\O^c,\O^d,\O^f]
\label{exe}
\ee
we readily obtain $\O^c=\O^d=1$ and $\O^f=-1$.
For the sake of brevity,
the summation symbol $\sum$
is omitted 
in the above expression (\ref{gtf}).

We find from 
the use of Eq. (8) through (\ref{exe})
that the ground state energy 
of the total system is
(see Appendix \ref{app1} for details)
\bn
E&=&Jsz\sum_{\vk}\left(\sqrt{1-\g_\vk^2}
+2D_{\vk,\vk}\right)-Js(s+1)zN\nc
&&+\f{Jz}{2}\f{2}{N}\sum_{\vko,\vkt,\vq}\g_\vq
\left(D_{\vkt,\vko-\vq}D_{\vko,\vkt+\vq}
-D_{\vko,\vko-\vq}D_{\vkt,\vkt+\vq}\right)
\label{get}
\en
where 
$D_{\vk,\vk'}=\f{1}{2}\left[
\left(1+\O^f\right)\left(\O^f+\bar{\O^f}\right)
\left(1+\bar{\O^f}\right)\right]_{\vk,\vk'}
\d_{\vk,\vk'}
=\lg f^\dag_{\vk'}f_\vk\rg
\d_{\vk,\vk'}$.
It is of note that
at half-filling 
$\svk D_{\vk,\vk}=0$.
We readily note from Eq. (\ref{get}) above that
the energy of free magnons 
at or near half-filling
is given by
$$
~~~~~~~~~~~~~~~~~~~~~~~~~~E_0=\svk\o_\vk^0-Js(s+1)zN,
~~~~~~~~~~~~~~~~~~~~~~~~~~~~~~~~~~~~~~~~~~(14.a)
$$
where $\o_\vk^0$ is the dispersion energy of magnons,
$$
~~~~~~~~~~~~~~~~~~~~~~~~~~~~~~~~\o_\vk^0=Jsz\sqrt{1-\g_\vk^2}.
~~~~~~~~~~~~~~~~~~~~~~~~~~~~~~~~~~~~~~~~~~~~~~~~~~~~~~~~(14.b)
$$
for a square lattice, 
the dispersion energy is simply,
$$~~~~~~~~~~~~~~~~~~~~~~~~~~~~~~~\o_\vk^0=2J\sqrt{1-\g_\vk^2}
~~~~~~~~~~~~~~~~~~~~~~~~~~~~~~~~~~~~~~~~~~~~~~~~~~~~~~~~~~(14.c)
$$
with $s=\f{1}{2}$.
Eq. (14.a) represents 
the familiar zero-point energy 
of the antiferromagnetic system
\cite{kit}.
The last term in Eq. (14.a) is 
the quantum contribution
which is essential 
for lowering energy.
\section{
Holstein-Primakoff Boson Approach
of $t-J$ Hamiltonian
by Functional Schr\"odinger Representation;
Antiferromagnetic Spin Polaron}
Based on the $t-J$ Hamiltonian,
we write the antiferromagnetic spin polaron Hamiltonian 
in momentum space\cite{Chang},
\setcounter{equation}{14}
\bn
H_{\mathrm AFP}&=&
\sum_\vq\o_\vq^0\left[1-\f{1}{2s}\f{2}{N}\sum_{\vq '}
\left(\f{\o_{\vq '}^0}{Jsz}-1\right)\right]
\left(c_\vq^\dag c_\vq +d_\vk^\dag d_\vk\right)
-\sqrt{2s}tz\sqrt{\f{2}{N}}\nc
&&\times\left(1-\f{1}{4s}\f{2}{N}
\sum_{\vko}\sinh^2\theta_{\vko}\right)
\left\{\sum_{\vk,\vq}f_{\vk}f_{\vk-\vq}^\dag
\left[g_{cc}(\vk,\vq)c_\vq^\dag
+g_{dd}(\vk,\vq)d_\vq\right]
+\mathrm{H.c.}\right\}
\nc
&&+Jsz\svk\left(\f{\o_\vk^0}{Jsz}-1
-2f_\vk^\dag f_\vk\right)
-Js^2zN.
\label{hafp}
\en
Here $\o_\vq^0=Jsz\sqrt{1-\g_\vq^2}$.
$t$ is the hopping integral.
$\theta_\vk$ is the mixing angle
defined by
$\tanh 2\theta_\vk = \g_\vk$.
$g_{cc}$ and $g_{dd}$ are the coupling functions given by
\bnn
~~~~~~~~~~~~~~~~~~~~~~~~~
g_{cc}(\vk,\vq)=\g_{\vk-\vq}\cosh\theta_\vq-\g_\vk\sinh\theta_\vq
~~~~~~~~~~~~~~~~~~~~~~~~~~~~~~~~~~~~~\mbox{(16.a)},\\  
~~~~~~~~~~~~~~~~~~~~~~~~~
g_{dd}(\vk,\vq)=-\g_{\vk-\vq}\sinh\theta_\vq+\g_\vk\cosh\theta_\vq
~~~~~~~~~~~~~~~~~~~~~~~~~~~~~~~~~~~~\mbox{(16.b)}.
\enn
The first and the last two terms 
in Eq. (\ref{hafp}) above
are identical to 
the terms that appear in Eq.(\ref{hz}).
It is reminded that 
the last two terms were not considered
in Ref.(4).

In order to properly
estimate
the influence of fermi-bose
(holon-magnon) coupling,
we now introduce 
the shifted wave functional\cite{Barnes}
associated with the scalar field,
\setcounter{equation}{16}
\be
\Psi_\O[\phi^c,\phi^d,u]
= N_\O\exp\left[-\f{1}{2}
\left(\Okc(\pkc-\pkzc)^2
+\Okd(\pkd-\pkzd)^2\right)
+\f{1}{2}u_\vk^\dag \O_{\vk, \vk '}^f
u_{\vk '}\right].
\label{sgtf}
\ee
The scalar field $\phi$ is shifted
by a constant $\phi_0$.
With the use of Eqs.(8) and (\ref{sgtf}) for Eq. (\ref{hafp}),
we obtain the ground state energy, 
$E=E[\phi_0^c,\phi_0^d;\O^c,\O^d,\O^f]$,
\bn
E&=&\lg H_{\mathrm AFP}\rg/\lg\O|\O\rg\nc
&=&\sum_\vq\o_\vq^0\left[1-\f{1}{2s}\f{2}{N}\sum_{\vq '}
\left(\f{\o_{\vq '}^0}{Jsz}-1\right)\right]
\left\{\left[\f{1}{4}\left(1-\Oqc^2\right)\Oqc^{-1}
-\f{1}{2}\left(1-\Oqc\right)\right]+\f{\pqzc^2}{2}
\right.\nc
&&\left.+\left[\f{1}{4}\left(1-\Oqd^2\right)\Oqd^{-1}
-\f{1}{2}\left(1-\Oqd\right)\right]+\f{\pqzd^2}{2}\right\}
+Jsz\svk\left(\f{\o_\vk^0}{Jsz}-1-2D_{\vk,\vk}\right)-Js^2zN\nc
&&-\sqrt{2s}tz\sqrt{\f{2}{N}}
\left(1-\f{1}{4s}\f{2}{N}
\sum_{\vko}\sinh^2\theta_{\vko}\right)
\sum_{\vk,\vq}\left(D_{\vk,\vk-\vq}+\d_{\vk,\vk-\vq}\right)
\left[g_{cc}(\vk,\vq)\f{1}{\sqrt{2}}\phi_{\vq 0}^c\right.\nc
&&\left.+g_{dd}(\vk,\vq)\f{1}{\sqrt{2}}\phi_{\vq 0}^d\right]
+{\mathrm H.c.}
\label{eafp}
\en
where 
$D_{\vk,\vl}=\f{1}{2}\left[(1+\O^f)(\O^f+\bar{\O}^f)^{-1}
(1+\bar{\O}^f)\right]_{\vk,\vl}$.
We take the variation of 
the above ground state energy 
with respect to 
the vacuum expectation value $\phi_0^c$,
\bn
&&\f{\d E}{\d \pqzc}\nc
&=&\svk\o_\vk^0\left[1-\f{1}{2s}\f{2}{N}\sum_{\vq '}
\left(\f{\o_{\vq '}^0}{Jsz}-1\right)\right]
\pkzc\d_{\vk,\vq}
-\sqrt{2s}tz\sqrt{\f{2}{N}}
\left(1-\f{1}{4s}\f{2}{N}
\sum_{\vq '}\sinh^2\theta_{\vq '}\right)\nc
&&\times\sum_{\vk,\vp}\left(D_{\vk,\vk-\vp}+\d_{\vk,\vk-\vp}\right)
g_{cc}(\vk,\vp)\f{1}{\sqrt{2}}
\d_{\vp,\vq}\nc
&=&\o_\vq^0\left[1-\f{1}{2s}\f{2}{N}\sum_{\vq '}
\left(\f{\o_{\vq '}^0}{Jsz}-1\right)\right]
\pqzc
-\sqrt{2s}tz\sqrt{\f{2}{N}}
\left(1-\f{1}{4s}\f{2}{N}
\sum_{\vq '}\sinh^2\theta_{\vq '}\right)\nc
&&\times\sum_{\vk,\vq}\left(D_{\vk,\vk-\vq}+\d_{\vk,\vk-\vq}\right)
g_{cc}(\vk,\vq)\f{1}{\sqrt{2}}\nc
&=&\o_\vq^0\left[1-\f{1}{2s}\f{2}{N}\sum_{\vq '}
\left(\f{\o_{\vq '}^0}{Jsz}-1\right)\right]
\pqzc
-\sqrt{2s}tz\sqrt{\f{2}{N}}
\left(1-\f{1}{4s}\f{2}{N}
\sum_{\vq '}\sinh^2\theta_{\vq '}\right)\nc
&&\times\f{1}{\sqrt{2}}\sum_{\vk,\vq}D_{\vk,\vk-\vq}g_{cc}(\vk,\vq)\nc
&=&0
\en

Thus from the above expression we readily find
the vacuum expectation value,
\be
\pqzc=\f{\sqrt{s}tz\sqrt{\f{2}{N}}
\left(1-\f{1}{4s}\f{2}{N}
\sum_{\vq '}\sinh^2\theta_{\vq '}\right)}
{\o_\vq^0\left[1-\f{1}{2s}\f{2}{N}\sum_{\vq '}
\left(\f{\o_{\vq '}^0}{Jsz}-1\right)\right]}
\sum_{\vk'}D_{\vk',\vk'-\vq}g_{cc}(\vk',\vq).
\label{pqzc}
\ee
We note that $D_{\vk',\vk'-\vq}=0$
for $|\vq|\neq 0$ and 
$\Oqc=\Oqd=1$.
Realizing the vacuum to vacuum transition,
we obtain the ground state energy 
from the insertion of
Eq. (\ref{pqzc}) into Eq. (\ref{eafp}),
\be
E=Jsz\sum_\vq\sqrt{1-\g_\vq^2}-Js(s+1)zN
+\svk\Sigma ^{\O}(\vk,\o)
\label{tjg}
\ee
where
\bn
\Sigma^{\O}(\vk,\o)
&=&-\f{st^2z^2\f{2}{N}\left[
\left(1-\f{1}{4s}\f{2}{N} 
\sum_{\vq '}\sinh^2\theta_{\vq '}\right)\right]^2} 
{\left[1-\f{1}{2s}\f{2}{N}\sum_\vq
\left(\f{\o_\vq^0}{Jsz}-1\right)\right]}
\nc
&&\times
\sum_\vq\left(D^2_{\vk,\vk-\vq}
g_{cc}^2(\vk,\vq)
+ D^2_{\vk,\vk-\vq}
g_{dd}^2(\vk,\vq)\right)/
\o_\vq^0.
\label{se}
\en
The first two terms in Eq. (\ref{tjg})
are identical to the expressions of
Eq. (14.a).
It is seen from Eqs. (\ref{tjg}) and (\ref{se})
that the ground state energy 
is lowered by
the self-energy of the holon(spinless hole)
which is contributed from
coupling to magnons.

\section{  
Slave-Boson Approach 
of $t-J$ Hamiltonian 
by Functional Schr\"{o}dinger Representation}
Here the slave-boson functional Schr\"{o}dinger representation
of the $t-J$ Hamiltonian 
will be discussed 
for the hole doped system 
of antiferromagnetically correlated electrons.
In this section emphasis is placed on 
a rigorous derivation 
of dispersion relation 
and its comparison with the Holstein-Primakoff theory
of Heisenberg Hamiltonian
in the limit of half-filling.
The $t-J$ Hamiltonian in the slave-boson representation 
is given by\cite{bza}
\[
H = -t\ds{\sum_{\lg ij\rg\s}}b_i b^\dag_j f^\dag_{i\s}f_{j\s}
-\f{J}{2}\ds{\sum_{\lg ij\rg\s}}
(f^\dag_{i\s}f^\dag_{j-\s}f_{j-\s} f_{i\s}
-f^\dag_{i\s}f^\dag_{j-\s}f_{j\s}f_{i-\s})
-\mu\ds{\sum_{i\s}}f^\dag
_{i\s}f_{i\s}.
~~~~~~~~~~~(23.a)
\]
Allowing a uniform hole doping rate,
$\d = b_ib_j^\dag$,
we rewrite
\[
H = -t\d\ds{\sum_{\lg ij\rg\s}}f^\dag_{i\s}f_{j\s}
-\f{J}{2}\ds{\sum_{\lg ij\rg\s}}
(f^\dag_{i\s}f^\dag_{j-\s}f_{j-\s} f_{i\s}
-f^\dag_{i\s}f^\dag_{j-\s}f_{j\s}f_{i-\s})
-\mu\ds{\sum_{i\s}}f^\dag
_{i\s}f_{i\s}.
~~~~~~~~~~~(23.b)
\]
Here $t$ is the hopping strength 
and $\mu$, the chemical  potential. 
$b_i (b^\dag_i)$ is the annihilation(creation) operator 
for a spinless boson and $f_{i\s}(f^\dag_{i\s})
$, the annihilation(creation) operator 
for a chargeless fermion with spin
$\s$ at site i.  

Following Floreanini and Jackiw\cite{fj}, 
the chargeless fermion field(spinon) operator $f$
can be written in terms of
the Grassmann field variables $u$
\setcounter{equation}{23}
\be
f_\a=\f{1}{\sqrt{2}}\left(u_\a+\f{\d}{\d u_\a}\right)
\label{fu}
\ee
where $\a$  represents $(i, \s)$.
We write Gaussian functional,
\be
|\O\rg\equiv \lg u|\Psi_\O\rg
= N_\O \exp (\f{1}{2}u_\a \O_{\a\b}u_\b)
\equiv N_\O \exp\f{1}{2}\tilde{u}\O u
\label{gf}
\ee
where $\O$ is the antisymmetric $2\times 2$ 
kernel matrix,
and $N_\O$, the normalization constant.
For brevity the summation sign 
over the lattice sites
is omitted 
in the above expression.
We now introduce 
Eqs. (\ref{fu}) and (\ref{gf}) 
into Eq. (24.b) to write the ground state energy,
\be
E=\f{\lg\O|H|\O\rg}{\lg\O|\O\rg}.
\label{eq4}
\ee
The resulting ground state energy is,
(for derivation,
see Appendix \ref{app2}),
\bn
E&=&-t\d\sum_{\lg ij\rg\s}D_{\s\s}(j,i)
-\mu\ds{\sum_{i\s}}D_{\s\s}(i,i)\nc
&&-\f{J}{2}\sum_{\lg ij\rg\s}
\left\{D_{\s\s}(i,i)D_{-\s-\s}(j,j)
+D_{\s\s}(j,i)D_{-\s-\s}(i,j)\right\},
\label{etj}
\en
where
\be
D_{\s\s^{'}}(i,j)=\f{1}{2}\Big\{\left[I+\O(i,j)\right]
  \left[\O(i,j)
+\overline{\O}(i,j)\right]^{-1}
 [I+\overline{\O}(i,j)]\left.\right\}_{\s\s^{'}}
\label{dss}
\ee
with $\s$ and $\s^{'}$, 
either the spin up or spin down state of electron.

We readily find from the inspection
of the chemical potential terms in Eqs. (24) and (\ref{etj})
that the number of electron $n_i$ at site $i$
and the local magnetization $m_i$ at site  $i$
are given by
\be
\left\{
\begin{array}{l}
n_i=D_{\uparrow\uparrow}(i,i)+D_{\downarrow\downarrow}(i,i)
=n_{i\uparrow}+n_{i\downarrow},\\
m_i
=D_{\uparrow\uparrow}(i,i)-D_{\downarrow\downarrow}(i,i)
= n_{i\uparrow}-n_{i\downarrow}.
\end{array}\right.
\label{nimi}
\ee
From Eq. (\ref{nimi}), we obtain
\be
\left\{\begin{array}{l}
D_{\uparrow\uparrow}(i,i)=(n_i+m_i)/2,\\
D_{\downarrow\downarrow}(i,i)=(n_i-m_i)/2.
\end{array}\right.
\label{di}
\ee
Insertion of Eq. (\ref{di}) 
into Eq. (\ref{etj}) leads to
\bn
E&=&-t\d\sum_{\lg ij\rg\s}D_{\s\s}(j,i)
-\f{J}{4}\sum_{\lg ij\rg\s}D_{\s\s}(i,i)\left(n_j-\s m_j\right)
-\f{J}{2}\sum_{\lg ij\rg\s}D_{\s\s}(j,i)D_{-\s-\s}(i,j)\nc
&&-\mu N(1-\d).
\label{emn}
\en

From the Fourier transform 
of expression (\ref{emn}) above
we obtain the following ground state energy
(see Appendix \ref{app2} for verification),
\be
E=\pm 2\svk\sqrt{\left[\left(4t\d\right)^2
-\left(Jm\right)^2\right]
\g_\vk^2
+J^2m^2}-\mu (1-\d)N
-2J(1-\d)^2N
\label{gse}
\ee
and the dispersion relation, 
\be
\o_\vk=\pm 2\sqrt{\left[\left(4t\d\right)^2
-\left(Jm\right)^2\right]
\g_\vk^2+J^2m^2}
\label{dr}
\ee
with 
$\g_\vk=\left[\cos(k_xa)+ \cos(k_ya)\right]/2$.\\
At half-filling, i.e.,$\d = 0$, Eq. (\ref{dr}) leads to
\be
\o_\vk=\pm 2Jm\sqrt{1-\g_\vk^2}.
\label{hfdr}
\ee
For the system of paramagnetic state, 
i.e., $m=0$, we obtain
\be
\o_\vk=\pm 8t\d\g_\vk.
\label{pdr}
\ee

For the case of 
vanishing hopping integral
or in the limit of large $J$, that is, 
$t/J\ll 1$,
we obtain
the dispersion energy
of spin waves(magnon)
from Eq. (\ref{dr}),
\be
\o_\vk= 2Jm\sqrt{1-\g_\vk^2}.
\label{hdr}
\ee
Realizing from Eq. (\ref{nimi}) 
that $m=1$ 
at half-filling,
we find that 
\be
\o_\vk= 2J\sqrt{1-\g_\vk^2},
\label{drsh}
\ee
We find from Eq. (\ref{drsh}) above
that in the limit of half-filling,
that is, $\d\ra 0$,
the slave-boson theory of FSP
leads to the identical dispersion relation
(Eq. (14.c)) of magnon obtained 
from the Holstein-Primakoff theory.

\section{summary}
In the present study
we showed the functional Schr\"odinger representations
of Holstein-Primakoff boson 
and slave-boson theories
for the Heisenberg Hamiltonian
and the $t-J$ Hamiltonian respectively.
From the use of
the functional Schr\"odinger picture(FSP) theory
of Holstein-Primakoff boson approach
for the  Heisenberg Hamiltonian
we obtained both the zero-point energy 
of free magnons
and the self-energy of holon(spinless holes).
The FSP 
of slave-boson theory was also introduced into
the $t-J$ Hamiltonian 
to derive the dispersion relation 
of magnon.
We find that in the 
limit of half-filling$(\d\ra 0)$
the dispersion relation 
derived from this approach
is identical to the one
obtained from the Holstein-Primakoff boson approach.
Improvement over the present approach
of functional Schr\"{o}dinger picture theory
is desirable to fully account for 
many-body effects(correlation effects) 
beyond the mean field level.
\acknowledgements
One (S.H.S.S.) of us greatly acknowledges 
the Korean Ministry of Education
(BSRI-1998) and the Center for Molecular Science
at Korea Advanced Institute of Science and Technology for
financial supports. 
We are grateful to Dr. I. E. Dikstein 
for helpful discussions.
\appendix
\section{
Evaluations of Gaussian Exponents, 
$\O^c$, $\O^d$ and $\O^f$ 
and Functional Schr\"odinger Representations 
of one- and two-body terms
for both bosons and fermions}
\label{app1}
In real space,
we define the ground state 
Gaussian functional of boson\cite{fj}
to be 
\be
\lg\phi|\Ob\rg \equiv\Psi_\O[\phi]
=N_\O ~e^{-\f{1}{2}\int\int\phi(\vx)\Ob(\vx -\vy)
\phi(\vy)~d\vx d\vy},
\label{twr}
\ee
or in an abbreviated form, 
we write 
$\Psi_\O[\phi] = e^{-\f{1}{2}\phi\Ob\phi}$.
$N_\O$ is the normalization constant.
The ground state functional
in momentum space
is written
\be
\Psi_\O[\tilde{\phi}]
=N_\O e^{-\f{1}{2}\int\phi^*_\vk\Okb\phi_\vk d\vk}
\label{twc}
\ee
or allowing discreteness 
in momentum $\vk$,
\be
\Psi_\O[\tilde{\phi}]
=N_\O\exp\left[-\f{1}{2{\cal N}}
\svk\tilde{\phi}_{-\vk}\Ob_\vk\tilde{\phi}_\vk\right]\nc
\ee
where ${\cal N}$ is the total number
of lattice sites.

Introducing the boson field $\phi$,
we define the boson operators in FSP,
\bn
c(\vx)&=&\f{1}{\sqrt{2}}
\left(\phi(\vx)+\f{\d}{\d\phi(\vx)}\right)\nc
c^\dag(\vx)&=&\f{1}{\sqrt{2}}
\left(\phi(\vx)-\f{\d}{\d\phi(\vx)}\right)
\en
which satisfy 
the commutation relation,
\be
[c(\vx),c^\dag(\vx')]=\d(\vx-\vx').
\label{crcc}
\ee
The field operators in momentum space
are then 
\bn
c(\vk)&=&\f{1}{\sqrt{2}}
\f{1}{(2\pi)^d}\int d\vx~e^{i\vk\cdot\vx}
\left(\phi(\vx)+\f{\d}{\d\phi(\vx)}\right)\nc
c^\dag(\vk)&=&\f{1}{\sqrt{2}}
\f{1}{(2\pi)^d}\int d\vx~e^{-i\vk\cdot\vx}
\left(\phi(\vx)-\f{\d}{\d\phi(\vx)}\right),
\en
and thus,
\bn
c(\vk)&=&\f{1}{\sqrt{2}}
\left(\tilde{\phi}(\vk)
+\f{\d}{\d\tilde{\phi}^*(\vk)}\right)\nc
c^\dag(\vk)&=&\f{1}{\sqrt{2}}
\left(\tilde{\phi}^*(\vk)
-\f{\d}{\d\tilde{\phi}(\vk)}\right),
\label{ckcdk}
\en
the use of which satisfies
the commutation relation,
$\left[c_\vk,c^\dag_{\vk'}\right]=\d_{\vk,\vk'}$.

Now we evaluate the vacuum state expectation
of one-body term ($\lg c_\vko^{\dag}c_\vkt\rg$)
in FSP.\\
Using the ground state Gaussian functional
\[
\Psi_\O[\tilde{\phi}]
=N_\O\exp\left[-\f{1}{2{\cal N}}
\sum_\vk\phi_{-\vp}\O_\vp\phi_\vp\right],
\]
we obtain
\bnn
\f{\d}{\d\phi^*_\vk}\Psi_\O[\tilde{\phi}]
&=&\f{\d}{\d\phi_{-\vk}}\Psi_\O[\tilde{\phi}]\nc
&=&-\f{1}{2}\sum_\vp\left(\Opb\d_{-\vk,-\vp}\phi_\vp
+\Opb\phi_{-\vp}\d_{-\vk,\vp}\right)\Psi_\O[\tilde{\phi}]\nc
&=&-\f{1}{2}\left(\Okb\phi_\vk
+\Omkb\phi_\vk\right)\Psi_\O[\tilde{\phi}].
\enn
By using the symmetric condition, 
$\Ob_\vk=\Ob_{-\vk}$,
we obtain
\bn
\f{\d}{\d\phi_\vk}\Psi_\O[\tilde{\phi}]
&=&-\Ob_\vk\phi_\vk\Psi_\O[\tilde{\phi}].
\en
Thus we find
\be
\lg 0|c_\vko^{\dag}c_\vkt|0\rg
\ra\f{1}{2}\left(1-\Oktb\right)\left(1+\Okob\right)
\int D\phi\Psi_\O^*[\tilde{\phi}]
\phi_\vko^{*}\phi_\vkt\Psi_\O[\tilde{\phi}]
-\f{1}{2}\left(1-\Oktb\right)\dkot
\label{occoi}
\ee
with
\bn
\int D\phi\Psi_\O^*[\tilde{\phi}]
\phi_\vko^{*}\phi_\vkt\Psi_\O[\tilde{\phi}]
&=&\int D\phi ~e^{-\phi^{*}\Ob\phi}\phi_\vko^{*}
\phi_\vkt~e^{-\phi^{*}\Ob\phi}\nc
&=&\int D\phi ~\phi_\vko^{*}\phi_\vkt
~e^{-2\phi^{*}\O^b\phi}\nc
&=&\lim_{J\rightarrow 0}
\f{\d^2}{\d J_\vko\d J_\vkt^{*}}
\left(\int D\phi ~e^{-2\phi^{*}\O^b\phi
+J^{*}\phi+J\phi^{*}}\right)\nc
&=&\lim_{J\rightarrow 0}
\f{\d^2}{\d J_\vko\d J_\vkt^{*}}
~e^{\f{1}{2}J^{*}{\O^b}^{-1}J}
\nc
&=&\lim_{J\rightarrow 0}\f{\d}{\d J_\vko}
\left(\f{1}{2}\Oktb^{-1}J_\vkt
e^{\f{1}{2}J^{*}{\Ob}^{-1}J}\right)\nc
&=&\lim_{J\rightarrow 0}\left(
\f{1}{2}\Oktb^{-1}\dkot 
+\f{1}{2}\Oktb^{-1}J_\vkt J_\vko 
e^{\f{1}{2}J^*\Ob^{-1}J}\right)\nc
&=&\f{1}{2}\Oktb^{-1}\dkot.
\label{oppo}
\en

By inserting Eq. (\ref{oppo}) into Eq. (\ref{occoi}), 
we obtain
\be
\lg 0|c_\vko^\dag c_\vkt|0\rg
\ra\f{1}{4}\left(1-\Oktb\right)\left(1+\Okob\right)
\Oktb^{-1}\dkot
-\f{1}{2}\left(1-\Oktb\right)\dkot . 
\label{ihob}
\ee
For $\vko=\vkt$, we get
\bn
\lg 0|c_\vk^\dag c_\vk|0\rg
&\ra&\f{1}{4}\left(1-\Okb\right)\left(1+\Okb\right)
\Okb^{-1}-\f{1}{2}\left(1-\Okb\right)\nc
&=&\f{1}{4}\left(\Okb^{-1}-\Okb\right)
-\f{1}{2}\left(1-\Okb\right)\nc
&=&\f{\Okb^{-1}+\Okb-2}{4}
\label{shob}
\en
In order to determine
the value of $\Ob$
for the energy of free boson
we now take the variation of 
the one-body energy term
($\lg c_\vk^\dag c_\vk\rg
\equiv\lg 0|c_\vk^\dag c_\vk|0\rg$)
with respect to $\O_\vk$,
\be
\f{\partial}{\partial\Okc}\lg c_{\vk}^{\dag}c_{\vk}\rg
=\f{1}{4}\left[-\f{1}{\left(\Okb\right)^2}+1\right]=0
\ee
to obtain $$\Okb=\pm 1.$$
Here $\Okb=-1$ is discarded
as the ground state Gaussian functional
cannot be defined.
Finally for $\vko=\vkt$,
Eq. (\ref{ihob}) leads to
\be
\lg 0|c_\vk^\dag c_\vk|0\rg=0 ~~ \mbox{for} ~~ \Okb=1
\ee
as expected.
In a similar manner
we obtain $\Okb=1$ 
for the two-body terms.

We note that for fermions
\[
f_a=\f{1}{\sqrt{2}}\left(u_a+\f{\d}{\d u_a^\dag}\right)
\]
and 
\[
\left\{f_a,f_b^\dag\right\}
=\d_{ab}
\]
with $i=a$ and $b$
to represent site indices.\\
We find
\be
\lg\O|\O\rg = \int DuDu^\dag
\exp[u^\dag(\O+\overline{\O})u] 
= {\det}^{\f{1}{2}}(\O+\overline{\O}).
\ee  
Realizing that
\bnn
f_b|0\rg 
&\ra& \f{1}{\sqrt{2}}\left(u_b+\f{\d}{\d u_b^\dag}\right)
\exp(u^\dag_c \O_{cd}u_d) \\
 &=&\f{1}{\sqrt{2}}(u_b+\d_{bc}\O_{cd}u_{d})
\exp(u^\dag_c \O_{cd}u_d) \\
 &=& (\d_{bd}+\O_{bd})u_d\exp(u^\dag_c\O_{cd}u_d)\\ 
 &=& (I+\O)_{bd}u_d\exp(u^\dag_c\O_{cd}u_d)
\enn
the expectation value of $\lg f^\dag f\rg$ is
\bn
\lg f_a^\dag f_b \rg 
&\ra&D_{ba}
\equiv\f{1}{2}\lg\O|\left(u_a^\dag+\f{\d}{\d u_a}\right)
\left(u_b+\f{\d}{\d u_b^\dag}\right)|\O\rg
/\lg\O|\O\rg 
\nc
&=& \f{1}{2}(I+\overline{\O})_{ea}(I+\O)_{bd}
\int DuDu^\dag u^\dag_e u_d\exp[u^\dag(\overline{\O}+\O)u]\nc
&=&\f{1}{2}(I+\overline{\O})_{ea}(I+\O)_{bd}
\f{\d}{\d\eta}\f{\d}{\d\bar{\eta}}Z_{0}[\eta,\bar{\eta}]|_{{\eta} 
= \bar{\eta}=0}\nc
&=& \f{1}{2}(I+\overline{\O})_{ea}(I+\O)_{bd}\f{\d}{\d \eta}
\f{\d}{\d \bar{\eta}}\exp[-\bar{\eta}(\O+\overline{\O})^{-1}\eta]
|_{\eta = \bar{\eta}=0}\nc
&=& \f{1}{2}(I+\overline{\O})_{ea}(I+\O)_{bd}
(\O+\overline{\O})^{-1}_{ed} \nc
&=& \f{1}{2}[(I+\O)(\O+\overline{\O})^{-1}(I+\overline\O)]_{ba}
\label{<ff>}
\en
where 
\be
Z_{0}[\eta,\bar{\eta}]= \int DuDu^\dag
\exp[\int(u^\dag Su+\bar{\eta} u+u^\dag\eta)]
=\det(S)\exp\left(-\bar{\eta}S^{-1}\eta\right)
\ee
with $S=\O+\overline{\O}$.
Similarly 
we find, for the two-body term, 
\bn
\lg f^\dag_a f^\dag_b f_c f_d\rg
&\ra& (I+\overline{\O})_{lb} (I+\overline{\O})_{ja}(I+\O)_{df}(I+\O)_{ch} 
    (\O+\overline{\O})^{-1}_{fn}(\O+\overline{\O})^{-1}_{hp}
    (\d_{pl}\d_{nj}-\d_{nl}\d_{pj})\nc 
&=& [(I+\O)_{df}(\O+\overline{\O})^{-1}_{fj}(I+\overline{\O})_{ja}]
    [(I+\O)_{ch}(\O+\overline{\O})^{-1}_{hl}(I+\overline{\O})_{lb}] \nc
&&  -[(I+\O)_{df}(\O+\overline{\O})^{-1}_{fl}(I+\overline{\O})_{lb}]
  [(I+\O)_{ch}(\O+\overline{\O})^{-1}_{hj}(I+\overline{\O})_{ja}] \nc
&=& D_{da}D_{cb}-D_{db}D_{ca}
\label{<ffff>}
\en
with $i=a,b,c$ and $d$, the site index.

With the use of Eq. (8), we obtain
in momentum space,
\bn
\lg H_0\rg
&=&\svk\o_{\vk}^0\left[1-\f{1}{2s}\f{2}{N}\sum_{\vq}
\left(\f{\o_{\vq}^0}{Jsz}-1\right)\right]
\left(\lg c_{\vk}^{\dag}c_{\vk}\rg+\lg d_{\vk}^{\dag}d_{\vk}\rg\right)
+Jsz\svk\left(\f{\o_\vk^0}{Jsz}-1-2\lg f_\vk^\dag f_\vk\rg\right)\nc
&&-Js^2zN\nc
&=&\svk\o_{\vk}^0\left[1-\f{1}{2s}\f{2}{N}\sum_{\vq}
\left(\f{\o_{\vq}^0}{Jsz}-1\right)\right]\left[
\f{1}{4}\left(1-\Okc\right)\left(1+\Okc\right)
\Okc^{-1}
-\f{1}{2}\left(1-\Okc\right)\right.\nc
&&\left.+\f{1}{4}\left(1-\Okd\right)\left(1+\Okd\right)
\Okd^{-1}
-\f{1}{2}\left(1-\Okd\right)
\right]
+Jsz\svk\left(\f{\o_\vk^0}{Jsz}-1-2D_{\vk,\vk}\right)\nc
&&-Js^2zN,
\label{ehz}
\en
\bn
\lg H_{m-m}\rg
&=&-\f{Jz}{4}\f{2}{N}\sum_{\vko,\vkt}\left[
\left(1-\f{\o_\vko^0\o_\vkt^0}{(Jsz)^2}\right)
\left(\lg c_{\vko}^{\dag}c_{\vko}c_{\vkt}^{\dag}c_{\vkt}\rg
+\lg d_{\vko}^{\dag}d_{\vko}d_{\vkt}^{\dag}d_{\vkt}\rg\right)
\right.\nc
&&\left.
+2\left(1+\f{\o_\vko^0\o_\vkt^0}{(Jsz)^2}\right)
\lg c_{\vko}^{\dag}c_{\vko}\rg\lg d_{\vkt}^{\dag}d_{\vkt}\rg\right]\nc
&=&-\f{Jz}{4}\f{2}{N}\sum_{\vko,\vkt}\left\{
\left(1-\f{\o_\vko^0\o_\vkt^0}{(Jsz)^2}\right)
\left[\f{1}{16}\left(1-\Oktc\right)\left(1+\Oktc\right)\right.\right.\nc 
&&\times\left(1-\Okoc\right)\left(1+\Okoc\right)
{\Oktc}^{-1}\left[{\Okoc}^{-1}+{\Oktc}^{-1}
\left(\dkot\right)^2\right]\nc
&&-\f{1}{8}\left(1-\Oktc\right)\left(1+\Oktc\right)
\left(1-\Okoc\right)\left({\Oktc}^{-1}
+\d_{\vko,\vkt}\Oktc^{-1}\d_{\vko,\vkt}\right)\nc
&&+\f{1}{8}\left(1-\Oktc\right)\left(1+\Oktc\right)
\d_{\vko,\vkt}\left(1+\Okoc\right)\Oktc^{-1}
\d_{\vko,\vkt}\nc
&&-\f{1}{4}\left(1-\Oktc\right)\left(1+\Oktc\right)
\left(\dkot\right)^2\nc
&&-\f{1}{8}\left(1-\Oktc\right)\left(1-\Okoc\right)
\left(1+\Okoc\right)\Okoc^{-1}\nc
&&+\f{1}{4}\left(1-\Oktc\right)\left(1-\Okoc\right)\nc
&&+\f{1}{16}\left(1-\Oktd\right)\left(1+\Oktd\right)
\left(1-\Okod\right)\left(1+\Okod\right)
{\Oktd}^{-1}\left[{\Okod}^{-1}+{\Oktd}^{-1}
\left(\dkot\right)^2\right]\nc
&&-\f{1}{8}\left(1-\Oktd\right)\left(1+\Oktd\right)
\left(1-\Okod\right)\left({\Oktd}^{-1}
+\d_{\vko,\vkt}\Oktd^{-1}\d_{\vko,\vkt}\right)\nc
&&+\f{1}{8}\left(1-\Oktd\right)\left(1+\Oktd\right)
\d_{\vko,\vkt}\left(1+\Okod\right)\Oktd^{-1}
\d_{\vko,\vkt}\nc
&&-\f{1}{4}\left(1-\Oktd\right)\left(1+\Oktd\right)
\left(\dkot\right)^2\nc
&&-\f{1}{8}\left(1-\Oktd\right)\left(1-\Okod\right)
\left(1+\Okod\right)\Okod^{-1}\nc
&&\left.+\f{1}{4}\left(1-\Oktd\right)\left(1-\Okod\right)\right]\nc
&&+2\left(1+\f{\o_\vko^0\o_\vkt^0}{(Jsz)^2}\right)
\left[\f{1}{4}\left(1-\Okc\right)\left(1+\Okc\right)
\Okc^{-1} 
-\f{1}{2}\left(1-\Okc\right)\right]\nc
&&\left.\times\left[\f{1}{4}\left(1-\Okd\right)\left(1+\Okd\right)
\Okd^{-1} 
-\f{1}{2}\left(1-\Okd\right)\right]\right\}, 
\en
\bn
\lg H_{m-h}\rg&=&-\f{Jz}{2}\f{2}{N}\sum_{\vko,\vkt}\f{\o_{\vk_2}^0}{Jsz}
\lg f_{\vko}^{\dag}f_{\vko}\rg\left(\lg c_{\vkt}^{\dag}c_{\vkt}\rg
+\lg d_{\vkt}^{\dag}c_{\vkt}\rg\right)\nc
&=&-\f{Jz}{2}\f{2}{N}\sum_{\vko,\vkt}\f{\o_{\vk_2}^0}{Jsz}
\f{1}{2}\left[(I+\O^f)(\O^f+\bar{\O}^f)^{-1}
(I+\bar{\O}^f)\right]_{\vko,\vko}\nc
&&\times\left[\f{1}{4}(1-\Oktc)(1+\Oktc)
\Oktc^{-1}-\f{1}{2}(1-\Oktc)\right.\nc
&&\left.+\f{1}{4}(1-\Oktd)(1+\Oktd)
\Oktd^{-1}-\f{1}{2}(1-\Oktd)\right]\nc
&=&-\f{Jz}{2}\f{2}{N}\sum_{\vko,\vkt}\f{\o_{\vk_2}^0}{Jsz}
\f{1}{2}\left[(I+\O^f)(\O^f+\bar{\O}^f)^{-1}
(I+\bar{\O}^f)\right]_{\vko,\vko}\nc
&&\times\left(\f{\Oktc^{-1}+\Oktc-2}{4}+\f{\Oktd^{-1}+\Oktd-2}{4}\right),
\en
and
\bn
H_{h-h}&=&-\f{Jz}{2}\f{2}{N}\sum_{\vko,\vkt}\g_\vq  
\lg f_{\vko-\vq}^{\dag}f_\vko f_{\vkt+\vq}^{\dag}f_\vkt\rg\nc
&=&-\f{Jz}{2}\f{2}{N}\sum_{\vko,\vkt}\g_\vq 
\lg\left[ f_{\vko-\vq}^{\dag}
\left(\d_{\vko,\vkt+\vq}-f_{\vkt+\vq}^{\dag}f_\vko\right)f_\vkt\right]\rg\nc
&=&-\f{Jz}{2}\f{2}{N}\sum_{\vko,\vkt}\g_\vq
\left(\lg f_{\vko-\vq}^{\dag}f_\vkt\rg\d_{\vko,\vkt+\vq}
-\lg f_{\vko-\vq}^{\dag}f_{\vkt+\vq}^{\dag}f_\vko f_\vkt\rg\right)\nc 
&=&-\f{Jz}{2}\f{2}{N}\sum_{\vq}\g_\vq
\left[\svk D_{\vk,\vk}
-\sum_{\vko,\vkt}\left(D_{\vkt,\vko-\vq}D_{\vko,\vkt+\vq}
-D_{\vko,\vko-\vq}D_{\vkt,\vkt+\vq}\right)\right]
\en
where
\be
D_{\vk,\vk '}=\f{1}{2}\left[(1+\O^f)(\O+\bar{\O}^f)(1+\bar{\O}^f)\right]
_{\vk,\vk '}
=\lg f_{\vk '}^\dag f_{\vk}\rg.
\ee

By noting that
\be
\svk\g_\vk=\svk\g_{-\vk}=0,
\ee
we obtain
\be
H_{h-h}=\f{Jz}{2}\f{2}{N}\sum_{\vko,\vkt,\vq}\g_\vq
\left(D_{\vkt,\vko-\vq}D_{\vko,\vkt+\vq}
-D_{\vko,\vko-\vq}D_{\vkt,\vkt+\vq}\right).
\ee
By taking the variation of the ground state energy
with respect to $\Okc$($\Okd$), 
we find that
$\Okc=\Okd=1$.
The resulting ground state energy is then
\bn
E&=&Jsz\sum_{\vk}\left(\sqrt{1-\g_\vk^2}
+2n_\vk^f\right)-Js(s+1)zN\nc
&&+\f{Jz}{2}\f{2}{N}\sum_{\vko,\vkt,\vq}\g_\vq
\left(D_{\vkt,\vko-\vq}D_{\vko,\vkt+\vq}
-D_{\vko,\vko-\vq}D_{\vkt,\vkt+\vq}\right)
\label{ge0}
\en
where $n_\vk^f\equiv\lg f^\dag_\vk f_\vk\rg=D_{\vk,\vk}$.
By taking the variation of 
$\lg H_0\rg$ in Eq. (\ref{ehz})
with respect to $\O^f_\vk$ we obtain
$\O^f_\vk=-1$, which leads to $D_{\vk,\vk'}=0$
at half-filling. 
Thus both the free hole energy term
and the hole-hole interaction energy 
(the second term in Eq.(\ref{ge0})) term vanish.

\section{the Dispersion Relation
of Magnons
from the $t-J$ Hamiltonian}
\label{app2}
From Eq. (23.b) the expectation value of the $t-J$ Hamiltonian
in slave-boson representation is
\bn
E&=&\lg H\rg\nc
 &=& -t\d\ds{\sum_{\lg ij\rg\s}} 
\lg f^\dag_{i\s}f_{j\s}\rg
-\f{J}{2}\ds{\sum_{\lg ij\rg\s}}
(\lg f^\dag_{i\s}f^\dag_{j-\s}f_{j-\s} f_{i\s}\rg
-\lg f^\dag_{i\s}f^\dag_{j-\s}f_{j\s}f_{i-\s}\rg)\nc
&&-\mu\ds{\sum_{i\s}}\lg f^\dag_{i\s}f_{i\s}\rg.
\label{<tj>}
\en
In the following we express
both $\lg f_a^\dag f_b\rg$
and $\lg f_a^\dag f_b f_c^\dag f_d\rg$
above in functional Schr\"odinger picture.
Using that
\bnn
f_b|0\rg 
&\ra& \f{1}{\sqrt{2}}\left(u_b+\f{\d}{\d\tu_b}\right)
\exp(\tu_c\O_{cd}u_d) \\
 &=&\f{1}{\sqrt{2}}\left(u_b+\d_{bc}\O_{cd}u_{d}\right)
\exp(\tu_c \O_{cd}u_d) \\
 &=& (\d_{bd}+\O_{bd})u_d\exp(\tu_c\O_{cd}u_d)\\ 
 &=& (I+\O)_{bd}u_d\exp(\tu_c\O_{cd}u_d),
\enn
we obtain
\bn
\lg f_a^\dag f_b\rg 
&=&\lg\O|f_a^\dag f_b|\O\rg / \lg\O|\O\rg\nc
&=& \f{1}{2}(I+\overline{\O})_{ea}(I+\O)_{bd}
\int DuD\tu \tu_e u_d\exp[\tu(\overline{\O}+\O)u]\nc
&=&\f{1}{2}(I+\overline{\O})_{ea}(I+\O)_{bd}
\f{\d}{\d\eta}\f{\d}{\d\bar{\eta}}Z_{0}[\eta,\bar{\eta}]|_{{\eta} 
=\bar{\eta}=0}\nc
&=& \f{1}{2}(I+\overline{\O})_{ea}(I+\O)_{bd}\f{\d}{\d \eta}
\f{\d}{\d \bar{\eta}}\exp[-\bar{\eta}(\O+\overline{\O})^{-1}\eta]
|_{\eta = \bar{\eta}=0}\nc
&=& \f{1}{2}(I+\overline{\O})_{ea}(I+\O)_{bd}
(\O+\overline{\O})^{-1}_{ed} \nc
&=& \f{1}{2}[(I+\O)(\O+\overline{\O})^{-1}(I+\overline\O)]_{ba}
\label{offo}
\en
where 
\be
Z_{0}[\eta,\bar{\eta}]= \int DuD\tu
\exp[\int(\tu Su+\bar{\eta} u+\tu\eta)]
=\det(S) \exp[-\bar{\eta}S^{-1}\eta].
\ee
For brevity we define
\be
D_{ba}=\f{1}{2}\left[
\left(I+\O\right)\left(\O+\overline{\O}\right)^{-1}
\left(I+\overline{\O}\right)\right]_{ba},
\ee
with $S=\O+\overline{\O}$,
and $\lg\O|\O\rg={\det}^{\f{1}{2}}
\left(\O+\overline{\O}\right)$.

We express
\bn
f_c f_d|0\rg 
&\ra&\f{1}{2}(u_c+\f{\d}{\d\tu_c})
(u_d+\f{\d}{\d\tu_d})
\exp(\tu_e \O_{ef}u_e) \nc
&=&\f{1}{2}(u_c+\f{\d}{\d\tu_c})
(I+\O)_{df}u_f\exp(\tu_g\O_{gh}u_h) \nc
&=&(I+\O)_{df}(u_c u_f+\f{\d}{\d \tu_c}u_f)
\exp(\tu_g\O_{gh}u_h)\nc
&=&(I+\O)_{df}(u_c u_f - u_f \O_{ch} u_h) e^{\tu\O u} \nc
&=&(I+\O)_{df}(I+\O)_{ch} u_h u_f e^{\tu\O u},
\en
and
\bn
\lg 0|f_a^\dag f_b^\dag f_c f_d|0\rg 
&=&\f{1}{4}(I+ \overline{\O})_{lb} (I+ \overline{\O})_{ja}
(I+\O)_{df} (I+\O)_{ch} \nc
& & \times \int DuD\tu e^{\tu\overline{\O} u}
\tu_j \tu_l u_h u_f e^{\tu\O u}   
\label{effff}
\en
where
\bn
&&\int DuD\tu ~e^{\tu\overline{\O} u}
\tu_j \tu_l u_h u_f e^{\tu\O u}   
= -\f{\d}{\d\eta_j}\f{\d}{\d\eta_l}\f{\d}{\d\bar{\eta_h}}
 \f{\d}{\d\bar{\eta_f}}Z_{0}[\eta,\bar{\eta}]|_{{\eta} 
=\bar{\eta}=0}\nc
&&= -{\det}^{\f{1}{2}}(\O+\overline{\O})\f{\d}{\d\eta_j}
\f{\d}{\d\eta_l}\f{\d}{\d\bar{\eta_h}}
\f{\d}{\d\bar{\eta_f}}
\exp[-\bar{\eta_m}(\O+\overline{\O})^{-1}_{mn}\eta_n]
|_{\eta = \bar{\eta}=0}\nc
&&={\det}^{\f{1}{2}}(\O+\overline{\O})\d_{fm}(\O+\overline{\O})^{-1}_{mn} 
\f{\d}{\d\eta_j}\f{\d}{\d\eta_l}\f{\d}{\d\bar{\eta_h}}
\eta_n \exp[-\bar{\eta_o}(\O+\overline{\O})^{-1}_{op}\eta_p]
|_{\eta=\bar{\eta}=0}\nc
&&=-{\det}^{\f{1}{2}}(\O+\overline{\O})\d_{fm}\d_{ho}
(\O+\overline{\O})^{-1}_{mn}(\O+\overline{\O})^{-1}_{op}
 \f{\d}{\d\eta_j}\f{\d}{\d\eta_l}\eta_n\eta_p 
\exp[-\bar{\eta_q}(\O+\overline{\O})^{-1}_{qr}\eta_r]
|_{\eta=\bar{\eta}=0} \nc
&&= -{\det}^{\f{1}{2}}(\O+\overline{\O})(\O+\overline{\O})^{-1}_{fn}
(\O+\overline{\O})^{-1}_{hp} 
\f{\d}{\d\eta_j}
(\d_{nl}\eta_p-\d_{pl}\eta_n)
e^{-\bar{\eta}(\O+\overline{\O})^{-1}\eta}
|_{\eta=\bar{\eta}=0}\nc
&&=-{\det}^{\f{1}{2}}(\O+\overline{\O})(\O+\overline{\O})^{-1}_{fn}
(\O+\overline{\O})^{-1}_{hp}(\d_{nl} \d_{pj} - \d_{pl} \d_{nj})
\label{euuuu}
\end{eqnarray}
Realizing that 
$\lg\O|\O\rg ={\det}^{\f{1}{2}}
(\O+\overline{\O})$\cite{fj}, 
we find from Eq. (\ref{effff})
and Eq. (\ref{euuuu}) that  
\bn
\lg f^\dag_a f^\dag_b f_c f_d\rg
&=& (I+\overline{\O})_{lb} (I+\overline{\O})_{ja}(I+\O)_{df}(I+\O)_{ch} 
    (\O+\overline{\O})^{-1}_{fn}(\O+\overline{\O})^{-1}_{hp}
    (\d_{pl}\d_{nj}-\d_{nl}\d_{pj})\nc 
&=& [(I+\O)_{df}(\O+\overline{\O})^{-1}_{fj}(I+\overline{\O})_{ja}]
    [(I+\O)_{ch}(\O+\overline{\O})^{-1}_{hl}(I+\overline{\O})_{lb}] \nc
&&  -[(I+\O)_{df}(\O+\overline{\O})^{-1}_{fl}(I+\overline{\O})_{lb}]
  [(I+\O)_{ch}(\O+\overline{\O})^{-1}_{hj}(I+\overline{\O})_{ja}] \nc
&=& D_{da}D_{cb}-D_{db}D_{ca}.
\label{offffo}
\en
The substitution of Eqs. (\ref{offo}) 
and (\ref{offffo}) into Eq. (\ref{<tj>})
leads to
\bn
E&=&-t\d\ds{\sum_{\lg ij\rg\s}}D_{\s\s}(j,i)
-\f{J}{2}\ds{\sum_{\lg ij\rg\s}}
\left\{\left[
D_{\s\s}(i,i)D_{-\s-\s}(j,j)-D_{-\s\s}(j,i)D_{\s-\s}(i,j)
\right]\right.\nc
&&\left.-\left[
D_{-\s\s}(i,i)D_{\s-\s}(j,j)-D_{\s\s}(j,i)D_{-\s-\s}(i,j)
\right]\right\}
-\mu\ds{\sum_{i\s}}D_{\s\s}(i,i).
\label{EDD}
\en
Allowing the global $SU(2)$ symmetry and thus
\be
D_{\s-\s}(i,j)=\lg f_{j-\s}^\dag f_{i\s}\rg =0,
\ee
Eq. (\ref{EDD}) is rewritten,
\bn
E&=&-t\d\ds{\sum_{\lg ij\rg\s}}D_{\s\s}(j,i)
-\f{J}{2}\ds{\sum_{\lg ij\rg\s}}
\left\{D_{\s\s}(i,i)D_{-\s-\s}(j,j)+D_{\s\s}(j,i)D_{-\s-\s}(i,j)\right\} \nc
&&-\mu\ds{\sum_{i\s}}D_{\s\s}(i, i).
\label{ed}
\en

From the inspection of the last term, 
we note that
the number of electron  $n_i$ at site $i$ 
and the local(site) magnetization  $m_i$ at site  $i$ 
are given by
\bn
\label{eq10}
n_i&=&D_{\ua\ua}(i,i)+D_{\da\da}(i,i)
=n_{i\ua}+n_{i\da},\\
\label{eq11}
m_i&=&D_{\ua\ua}(i,i)-D_{\da\da}(i,i)
= n_{i\ua}-n_{i\da}.
\en
From Eqs. (\ref{eq10}) and (\ref{eq11}), we obtain
\bn
D_{\ua\ua}(i,i)&=&\f{1}{2}(n_i+m_i),\nc
D_{\da\da}(i,i)&=&\f{1}{2}(n_i-m_i),\nc
\sum_\s D_{\s\s}(i,i)&=&D_{\ua\ua}(i,i)
+D_{\da\da}(i,i)=n_{i} \nc
\sum_\s D_{\s\s}(i,i)D_{-\s-\s}(j,j)&=&D_{\ua\ua}(i,i)
D_{\da\da}(j,j)+D_{\da\da}(i,i)
D_{\ua\ua}(j,j)\nc
&=&D_{\ua\ua}(i,i)\f{\left(n_{j}-m_{j}\right)}{2}
+D_{\da\da}(i,i)\f{\left(n_{j}+m_{j}\right)}{2} \nc
&=&\f{1}{2}\sum_\s D_{\s\s}(i,i)(n_{j}-\s m_{j})
\label{dnm}
\en
with $\s=1(-1)$ for up(down) spin.
Using Eq. (\ref{dnm}),
we rewrite Eq. (\ref{ed}),
\bn
E&=&-t\d\ds{\sum_{\lg ij\rg\s}}D_{\s\s}(j,i)
-\f{J}{4}\sum_{\lg ij\rg}D_{\s\s}(i,i)(n_{j}-\s m_{j})
-\f{J}{2}\ds{\sum_{\lg ij\rg\s}}D_{\s\s}(j, i)D_{-\s-\s}(i,j)\nc
&&-\mu \sum_{i}n_{i}.
\label{enm'}
\en

Allowing uniform hole doping and thus $n=n_i=1-\d$,
we obtain the total number of electrons
\be
\sum_{i}n_{i}=\sum_{i}(1-\d)=N(1-\d)
\label{sni}
\ee
and
\be
\sum_{\lg ij\rg}\left[\sum_\s D_{\s\s}(i,i)\right]n_j
=\sum_{\lg ij\rg}n_in_j=\sum_{\lg ij\rg}\left(1-\d\right)^2
=4N\left(1-\d\right)^2
\label{sninj}
\ee
With the use of Eqs. (\ref{sni}) and (\ref{sninj}),
Eq. (\ref{enm'}) leads to
\bn
E&=&-t\d\ds{\sum_{\lg ij\rg\s}}D_{\s\s}(j,i)
+\f{J}{4}\sum_{\lg ij\rg}D_{\s\s}(i,i)\s m_j
-\f{J}{2}\ds{\sum_{\lg ij\rg\s}}D_{\s\s}(j,i)D_{-\s-\s}(i,j)\nc
&&-JN\left(1-\d\right)^2-\mu N(1-\d).
\label{emdi}
\en

We define
\bn
n(i,j)&=&D_{\ua\ua}(i,j)+D_{\da\da}(i,j), \nc
m(i,j)&=&D_{\ua\ua}(i,j)-D_{\da\da}(i,j).
\label{eq18}
\en
Then we have
\bn
D_{\ua\ua}(i,j)&=&\f{1}{2}[n(i,j)+m(i,j)], \nc
D_{\da\da}(i,j)&=&\f{1}{2}[n(i,j)-m(i,j)].
\label{dijnm}
\en
Using Eq. (\ref{dijnm}), 
we write the third term in Eq. (\ref{emdi}),
\bn
\sum_{\lg ij\rg\s}D_{\s\s}(j,i)D_{-\s-\s}(i,j)&=&
\sum_{\lg ij\rg}[D_{\ua\ua}(j,i)D_{\da\da}(i,j)
+D_{\da\da}(j,i)D_{\ua\ua}(i,j)] \nc
&=&\sum_{\lg ij\rg}\left[D_{\ua\ua}(j,i)\cdot 
\f{\{n(i,j)-m(i,j)\}}{2}
\right.\nc
&&\left.+D_{\da\da}(j,i)\cdot
\f{\{n(i,j)+m(i,j)\}}{2}\right]\nc
&=&\f{1}{2}\sum_{\lg ij\rg\s}D_{\s\s}(j,i)[n(i,j)-\s m(i,j)],
\label{ddnm}
\en
The substitution of Eq. (\ref{ddnm}) 
into Eq. (\ref{emdi}) leads to 
\bn
E&=&-t\d\sum_{\lg ij\rg\s}D_{\s\s}(j,i) 
+\f{J}{4}\sum_{\lg ij\rg\s}D_{\s\s}(i,i)\s m_j
-\f{J}{4}\sum_{\lg ij\rg\s}D_{\s\s}(j,i)\left[n(i,j)-\s m(i,j)\right]\nc
&&-JN\left(1-\d\right)^2-\mu N(1-\d).
\label{emdf}
\en

We note that 
the two-dimensional staggered magnetization 
is represented by $m_i=m~e^{i{\mathbf Q}\cdot{\mathbf i}}$ 
with ${\mathbf Q}=(\pi,\pi)$ 
and ${\mathbf i}=(i_x,i_y)$.
Thus Eq. (\ref{emdf}) leads to
\bn
E&=&-t\d\sum_{\lg ij\rg\s}D_{\s\s}(j,i) 
+\f{Jm}{4}\sum_{\lg ij\rg\s}\s D_{\s\s}(i,i)
e^{i{\mathbf Q}\cdot{\mathbf j}}
-\f{J}{4}\sum_{\lg ij\rg\s}D_{\s\s}(j,i)
\left[n(i,j)-\s m(i,j)\right]\nc
&&-\mu N(1-\d)-JN(1-\d)^2
\label{emdf'}
\en
where
\bn
\sum_{\s}D_{\s\s}(j,i)
&=&D_{\ua\ua}(j,i)+D_{\da\da}(j,i)
=n(j,i),\nc
\sum_{\lg ij\rg\s}D_{\s\s}(j,i)n(i,j)
&=&\sum_{\lg ij\rg}n(j,i)n(i,j).
\label{eq22}
\en
Introducing 
\be
\sum_{\lg ij\rg\s}D_{\s\s}(j,i)n(i,j) 
\simeq \sum_{\lg ij\rg}n_{i}n_{j}=4N(1-\d)^{2}
\label{eq23}
\ee
and
\be
m(i,j)\simeq m_i=m e^{i{\bf Q}\cdot{\bf i}},
\label{eq25}
\ee
we obtain from Eq. (\ref{emdf'}),
\bn
E&=&-t\d\sum_{\lg ij\rg\s}D_{\s\s}(j,i)
+\f{Jm}{4}\sum_{\lg ij\rg\s}\s D_{\s\s}(i,i)
e^{i{\mathbf Q}\cdot{\mathbf j}}
+\f{Jm}{4}\sum_{\lg ij\rg\s}\s D_{\s\s}(j,i)
e^{i{\mathbf Q}\cdot{\mathbf i}}\nc
&&-\mu N(1-\d)-2JN(1-\d)^2\nc
&=&\sum_{\lg ij\rg\s}\left\{D_{\s\s}(j,i)
\left[-t\d+\f{Jm}{4}\s 
e^{i{\mathbf Q}\cdot{\mathbf i}}\right]
+\f{Jm}{4}\s D_{\s\s}(i,i)
e^{i{\mathbf Q}\cdot{\mathbf j}}\right\}\nc
&&-\mu N(1-\d)-2JN(1-\d)^2.
\label{enmq}
\en

In momentum space, we note that 
for the case of square lattice,
\be
D_{\s\s}(j,i) = \f{1}{(2\pi)^{2}}\sum_{\vk,\vk'}
d_{\s\s}(\vk,\vk')e^{i\vk\cdot\vj}e^{-i\vk'\cdot\vi},
\label{eq27}
\ee
and thus
\bn
\sum_{\lg ij\rg}D_{\s\s}(j,i) &=&
\f{1}{(2 \pi)^{2}}\sum_{\lg ij\rg}\sum_{\vk,\vk'}
d_{\s\s}(\vk,\vk')e^{i\vk\cdot\vj}e^{-i\vk'\cdot\vi}\nc
&=&\f{1}{(2\pi)^{2}}\sum_{\vk,\vk'}d_{\s\s}(\vk,\vk')
\sum_{\lg ij\rg}e^{i\vk\cdot\vj}e^{-i\vk'\cdot\vi}.
\label{dkk}
\en
Using
\bn
\sum_{\lg ij\rg}e^{i\vk\cdot\vj}e^{-i\vk'\cdot\vi}
&=&\sum_{\vi,\va}e^{i\vk\cdot(\vi+\va)}
e^{-i\vk'\cdot\vi} \nc
&=&\sum_\vi e^{i(\vk-\vk')\cdot\vi}
\sum_\va e^{i\vk\cdot\va}\nc 
&=&(2 \pi)^2 \d(\vk-\vk') \times 2(\cos k_xa+\cos k_ya),
\label{eq29}
\en
with $\va$, the four nearest neighbour sites
around site $i$,
we obtain from Eq. (\ref{dkk}),
\bn
\sum_{\lg ij\rg}D_{\s\s}(j,i) &=&
\sum_{\vk,\vk'}d_{\s\s}(\vk,\vk')\d(\vk-\vk')
\times 2(\cos k_xa + \cos k_ya) \nc
&=&\svk d_{\s\s}(\vk,\vk)2(\cos k_xa+\cos k_ya)
\label{dkg}
\en

Noting that
\bn
\sum_{\lg ij\rg}D_{\s\s}(j,i)e^{i\vQ\cdot\vi}
&=&\f{1}{(2 \pi)^{2}}\sum_{\lg ij\rg}\sum_{\vk,\vk'}
d_{\s\s}(\vk,\vk')e^{i\vk\cdot\vj}e^{-i\vk'\cdot\vi}
e^{i\vQ\cdot\vi} \nc
&=&\f{1}{(2 \pi)^{2}}\sum_{\vk,\vk'}d_{\s\s}(\vk,\vk')
\sum_{\vi,\va}e^{i\vk\cdot(\vi + \va)}
e^{-i\vk'\cdot\vi}e^{i\vQ\cdot\vi} \nc
&=&\f{1}{(2 \pi)^{2}}\sum_{\vk,\vk'}d_{\s\s}(\vk,\vk')
\sum_\vi e^{i(\vk-\vk'+\vQ)\cdot\vi}
\sum_\va e^{i\vk\cdot\va} \nc
&=&\sum_{\vk,\vk'}d_{\s\s}(\vk,\vk')\d (\vk-\vk'+\vQ)
2(\cos k_xa + \cos k_ya)  \nc
&=&\sum_\vk d_{\s\s}(\vk,\vk+\vQ)2(\cos k_xa + \cos k_ya),
\label{dkqg}
\en
and
\bn
\sum_{\lg ij\rg}D_{\s\s}(i,i)e^{i\vQ\cdot\vj}
&=&\f{1}{(2\pi)^2}\sum_{\vi,\va}\sum_{\vk ,\vk '}
d_{\s\s}(\vk,\vk')e^{i\vk\cdot\vi}
e^{-i\vk '\cdot\vi}
e^{i\vQ\cdot\left(\vi+\va\right)}\nc
&=&\f{1}{(2\pi)^2}\sum_{\vk ,\vk '}d_{\s\s}(\vk,\vk')
\sum_\vi e^{i\left(\vk+\vQ-\vk '\right)\cdot\vi}
\sum_\va e^{i\vQ\cdot\va}\nc
&=&-4\sum_\vk d_{\s\s}(\vk,\vk+\vQ),
\label{dkq})
\en
we obtain,
with the use of Eq. (\ref{dkg}) above,
\bn
E&=&-t\d\sum_{\vk,\s}2(\cos k_xa+\cos k_ya)d_{\s\s}(\vk,\vk)
+\f{Jm}{2}\sum_{\vk,\s}\s(\cos k_xa+\cos k_ya)
d_{\s\s}(\vk,\vk+\vQ)\nc
&&-Jm\sum_{\vk,\s}\s d_{\s\s}(\vk,\vk+\vQ)
-\mu N(1-\d)-2JN(1-\d)^2. 
\label{eq32}
\en
Introducing $\g_\vk =\left(\cos k_xa+\cos k_ya\right)/2$,
the above equation is rewritten
\bn
E&=&\sum_{\vk,\s}\left[-4t\d\g_\vk d_{\s\s}(\vk,\vk)
+Jm\s\left(\g_\vk-1\right)d_{\s\s}(\vk,\vk+\vQ)\right]\nc
&&-\mu N(1-\d)-2JN(1-\d)^2.
\label{eq33}
\en

We take $\vk$
-sum in the reduced Brillouin zone
i.e., half the first Brillouin zone,
to write
\bn
\sum_{\vk,\s}\g_\vk d_{\s\s}(\vk,\vk)
&=&\sum_{\vk,\s}^{'}\left[\g_\vk d_{\s\s}(\vk,\vk) 
+\g_{\vk+\vQ}d_{\s\s}(\vk+\vQ,\vk+\vQ)\right]
\nc &=& \g_\vk\sum_{\vk,\s}^{'}\left [d_{\s\s}(\vk,\vk)
-d_{\s\s}(\vk+\vQ,\vk+\vQ)\right],
\label{eq34}
\en
and
\bn
\sum_{\vk,\s}\left(\g_\vk-1\right) d_{\s\s}(\vk,\vk+\vQ) 
&=&\sum_{\vk,\s}^{'}\left[\left(\g_\vk-1\right)
d_{\s\s}(\vk,\vk+\vQ)
+\left(\g_{\vk+\vQ}-1\right)
d_{\s\s}(\vk+\vQ,\vk+2\vQ)\right ]
\nc &=&\sum_{\vk,\s}\left[\left(\g_\vk-1\right)
d_{\s\s}(\vk,\vk+\vQ) 
-\left(\g_\vk+1\right)
d_{\s\s}(\vk+\vQ,\vk)\right].
\label{eq35}
\en
where we used the nesting condition, 
$\g_{\vk+\vQ}=-\g_\vk$.
The symbol $'$ indicates 
momentum summation 
in the reduced Brillouine zone.
By applying (\ref{eq34}) and (\ref{eq35}) to (\ref{eq33}), 
we obtain the ground state energy
of the two-dimensional hole-doped systems,
\bn
E&=&\sum_{\vk,\s}^{'}\left [-4t\d\g_\vk d_{\s\s}(\vk,\vk)
+Jm\s\left(\g_\vk-1\right)d_{\s\s}(\vk,\vk+\vQ)\right.\nc
&&\left.-Jm\s\left(\g_\vk+1\right)d_{\s\s}(\vk+\vQ,\vk)
 +4t\d\g_\vk d_{\s\s}(\vk+\vQ,\vk+\vQ)\right]\nc
&&-\mu N(1-\d)-2JN(1-\d)^2.
\en

Realizing the equivalence between
$d_{\s\s}(\vk,\vk')$ and $f_{\vk'}^\dag f_{\vk'}$ 
from the inspection of Eq. (23.b) and Eq. (\ref{etj}),
we consider
the transformation matrix in
\[
\left[\begin{array}{cc}
f_\vk & f_{\vk+\vQ}^\dag
\end{array}\right]
\left[\begin{array}{cc}
-4t\d\g_\vk & Jm\s\left(\g_\vk-1\right) \\
 -Jm\s\left(\g_\vk+1\right) & 4t\d\g_\vk
\end{array}\right]
\left[\begin{array}{c}
f_\vk^\dag \\
f_{\vk+\vQ}
\end{array}\right],
\]

From the determinant of the above matrix,
we obtain
\be
\epsilon^2-\left(4t\d\g_\vk\right)^2
+J^2m^2(\g_\vk^2-1) = 0.
\label{eq36}
\ee
The dispersion energy is
then
\be
\epsilon=\pm\sqrt{\left(4t\d\g_\vk\right)^2 
- J^2m^2\left(\g_\vk^2-1\right)}.
\label{eq37}
\ee
Considering summation over
the up and down spins, we write
the dispersion energy,
\bn
\o_\vk&=&\sum_{\s}\sqrt{\left[\left(4t\d\right)^2
-\left(Jm\right)^2\right]\g_\vk^2
+J^2m^2}\nc
&=&2\sqrt{\left[\left(4t\d\right)^2
-\left(Jm\right)^2\right]\g_\vk^2
+J^2m^2}.
\en

The ground state energy of 
the hole-doped two-dimensional antiferromagnet
is then
\be
E_\vk=\pm 2\svk\sqrt{\left[\left(4t\d\right)^2
-\left(Jm\right)^2\right]\g_\vk^2
+J^2m^2}-\mu(1-\d)-2J(1-\d)^2.
\label{eq39}
\ee
For the paramagnetic states, $m=0$, we obtain
\be
\o_\vk =\pm 4t\d\g_\vk.
\label{eq41}
\ee
For the undoped antiferromagnet, that is, $\d = 0$,
the dispersion energy 
of the antiferromagnetic magnon
is
\be
\o_\vk=2Jm\sqrt{1-\g_\vk^2}.
\label{eq42}
\ee

\end{document}